\documentclass[twocolumn,superscriptaddress,float,aps]{revtex4-2}
\usepackage{graphicx,amsfonts,amssymb,amsmath,hyperref,hypcap,enumerate}
\usepackage{color}
\usepackage{bm}
\usepackage{multirow}
\usepackage{makecell}
\usepackage{color}
\usepackage{hyperref}
\hypersetup{
colorlinks = true,
linkcolor = [rgb]{0.70,0.13,0.13},
citecolor = [rgb]{0.13,0.55,0.13},
urlcolor = [rgb]{0.25, 0.41, 0.88}}
\usepackage{makecell}

\newcommand{\dd}{\mathrm{d}}

\newcommand{\ii}{\mathrm{i}}
\newcommand{\e}{\mathrm{e}}

\newcommand{\U}{\mathrm{U}}

\newcommand{\dsZ}{\mathbb{Z}}

\newcommand{\scL}{\mathcal{L}}

\newcommand{\eq}[1]{\begin{equation}#1\end{equation}}
\newcommand{\eqs}[1]{\begin{equation}\begin{split}#1\end{split}\end{equation}}
\newcommand{\eqnref}[1]{Eq.\,\eqref{#1}}
\newcommand{\figref}[1]{Fig.\,\ref{#1}}
\newcommand{\tabref}[1]{Tab.\,\ref{#1}}

\begin{document}
\title{Symmetric Mass Generation in the 
1+1 Dimensional Chiral Fermion 3-4-5-0 Model}

\author{Meng Zeng}
\affiliation{Department of Physics, University of California San Diego, La Jolla, California 92093, USA}
\author{Zheng Zhu}
\affiliation{Kavli Institute for Theoretical Sciences, University of Chinese Academy of Sciences, Beijing 100190, China}
\affiliation{CAS Center for Excellence in Topological Quantum Computation, University of Chinese Academy of Sciences, Beijing 100190, China}
\author{Juven Wang}
\affiliation{Center of Mathematical Sciences and Applications, Harvard University, MA 02138, USA}
\author{Yi-Zhuang You}
\affiliation{Department of Physics, University of California San Diego, La Jolla, California 92093, USA}

\begin{abstract} Lattice regularization of chiral fermions has been a long-standing problem in physics. In this work, we present the density matrix renormalization group (DMRG) simulation of the 3-4-5-0 model of (1+1)D chiral fermions with an anomaly-free chiral U(1) symmetry, which contains two left-moving and two right-moving fermions carrying U(1) charges 3,4 and 5,0, respectively. Following the Wang-Wen chiral fermion model, we realize the chiral fermions and their mirror partners on the opposite boundaries of a thin strip of (2+1)D lattice model of multi-layer Chern insulator, whose finite-width implies the quantum system is effectively (1+1)D. By introducing carefully designed two sets of six-fermion local interactions to the mirror sector only, we demonstrate that the mirror fermions can be gapped out by the interaction above a critical strength without breaking the chiral U(1) symmetry, via the symmetric mass generation (SMG) mechanism. We show that the interaction-driven gapping transition is in the Berezinskii-Kosterlitz-Thouless (BKT) universality class. We determine the evolution of Luttinger parameters before the transition, which confirms that the transition happens exactly at the point when the interaction term becomes marginal. As the mirror sector is gapped after the transition, we check that the fermions in the light chiral fermion sector remain gapless, which provides the desired lattice regularization of chiral fermions.
\end{abstract}

\maketitle

\emph{Introduction.} --- It has been a long-standing issue to regularize chiral gauge theories (e.g.,~the weak interaction in the Standard Model) on the lattice due to the Nielsen-Ninomiya no-go theorem \cite{nielsen1981no}, which asserts that any free  fermion lattice model in even-dimensional spacetime with locally-realized chiral symmetry will necessarily give rise to equal numbers of left-handed and right-handed fermion fields at low energy, hence rendering the theory vector-like. 
Over the past few decades, much effort \cite{ginsparg1982remnant,Swift1984The-electroweak,Smit1986Fermions,Banks1992Decoupling,kaplan1992method,shamir1993chiral} has been devoted to circumvent the fermion doubling problem by lifting different assumptions of the no-go theorem. 
 
In particular, the no-go theorem assumes the fermion theory to be infrared free, i.e.,~fermion interactions, if there are any, must be perturbatively irrelevant under the renormalization group (RG) flow. Lifting this assumption by introducing non-perturbative (strong enough) fermion interactions could potentially circumvent the problem. Efforts along this line are generally referred to as the \textit{mirror fermion} approach, which dates back to Eichten and Preskill \cite{eichten1986chiral}. The basic idea is to start with a vector-like theory containing both chiral fermions and their mirror fermion partners, which can be put on a lattice without any issue. Then one attempts to generate a mass gap in the mirror sector by introducing interactions among mirror fermions, such that the remaining light (chiral fermion) sector survives in the low-energy spectrum, providing the basis for lattice realizations of chiral gauge theories. However, early numerical tests \cite{Bock1990Unquenched,Bock1992Fermion,Bock1993Fermion-Higgs,Bock1994Staggered,Hasenfratz1990Phase,Hasenfratz1991The-equivalence,Lee1990Study,Montvay1992Mirror,Golterman1993Absence, chen2013decoupling} appeared to invalidate the mirror fermion approach, as strong fermion interactions typically result in the condensation of fermion bilinear mass at low energy, which spontaneously breaks the chiral symmetry and gaps out the light sector together with the mirror sector.

In recent years, a series of developments \cite{Fidkowski:2010bf,Fidkowski:2011dd,Turner2011Topological,Ryu:2012ph,Qi:2013qe,Yao:2013yg, Wen1305.1045, Wang2013Non-Perturbative, Wang2014Interacting,Gu:2014tw,Metlitski2014Interaction,You:2014ho,Witten:2016yb,
Kikukawa1710.11101,
wang2019solution, WangWen1809.11171} in the many-body quantum matter community have significantly deepened our understanding. It is realized that in order to gap out the mirror sector by interactions without breaking the chiral symmetry, two conditions must be satisfied: (i) the mirror fermions must be \emph{anomaly free} under the full spacetime-internal symmetry, (ii) the interaction must be appropriately designed to satisfy certain \emph{consistent  gapping conditions} \cite{Wang2013Non-Perturbative, wang2019solution}.  Along this line, recent numerical studies \cite{Ayyar2015Massive,Slagle:2015lo,Catterall:2016sw,Ayyar:2016fi,Catterall:2016nh,Ayyar:2016tg,Ayyar:2016ph,He:2016qy,DeMarco2017A-Novel,Ayyar2017Generating,Schaich2018Phases,Catterall2018Topology,Butt2018Four,Butt2018SO4-invariant,Catterall2020Exotic,Catterall2021Chiral,Butt2021Symmetric} have successfully demonstrated examples of interaction-driven fermion mass generation without spontaneous symmetry breaking in various spacetime dimensions. The phenomenon is known as the \emph{symmetric mass generation} (SMG) \cite{You2018Symmetric,You2018From,Razamat2021Gapped,Tong2021Comments,Butt2021Anomalies}. Therefore, solving the chiral fermion problem boils down to achieving the SMG for mirror fermions in even spacetime dimensions.

Nevertheless, most numerical works realizing SMG in even spacetime dimensions have been focused on vector-like lattice models \cite{Catterall:2016nh,Ayyar:2016tg,Ayyar:2016ph,Ayyar2017Generating,Schaich2018Phases,Butt2018SO4-invariant,Catterall2020Exotic,Butt2021Symmetric}, which still have some distance from the goal of regularizing chiral fermions. Recently, Catterall \cite{Catterall2021Chiral} 
studied the SMG of a chiral fermion lattice model with a chiral discrete $\dsZ_4$ symmetry. 
In this work, we demonstrate the SMG in the 3-4-5-0 model of (1+1)D chiral fermions that cancels the $\dsZ$-class \emph{perturbative} local anomaly of the chiral continuous $\U(1)$ symmetry, which is closer to the situation of perturbative chiral anomaly cancellation 
in the (3+1)D Standard Model (such as the chiral U(1)$_Y$ electroweak hypercharge). 
We propose a lattice model of interacting fermions, and investigate the model using the density matrix renormalization group (DMRG) numerical method \cite{White1992Density,Schollwock2005DMRG}. Our numerical results provide clear evidence for the SMG in the mirror sector, successfully achieving our goal of regularizing chiral fermions in the 3-4-5-0 model on a lattice.  

\emph{The 3-4-5-0 Model.} --- The 3-4-5-0 model describes four gapless complex fermions in (1+1)D,
\eq{\label{eq:3450}
S=\int\dd t\;\dd x\;\sum_{I=1}^{4}\psi_I^\dagger (\ii \partial_t+\ii v_I\partial_x)\psi_I,}
with two left-moving modes $\psi_1,\psi_2$ (of $v_1=v_2=+1$) and two right-moving modes $\psi_3,\psi_4$ (of $v_3=v_4=-1$). The fermions are charged under a chiral $\U(1)$ symmetry: $\psi_I\to\e^{\ii q_I \theta}\psi_I$, with the charge assignment $(q_1,q_2,q_3,q_4)=(3,4,5,0)$ (hence the name ``3-4-5-0''). This seemly peculiar charge assignment is designed to cancel the $\U(1)$ symmetry's 't Hooft anomaly, which is an $\dsZ$-class perturbative local anomaly. The anomaly index is given by $\sum_{I}v_I q_I^2=3^2+4^4-5^2-0^2=0$, which vanishes for the charge assignment of the 3-4-5-0 model. The model is also free of the gravitational anomaly. As the field theory is anomaly-free, it should admit a lattice regularization in (1+1)D spacetime.

\begin{figure}[]
\includegraphics[width=0.9\linewidth]{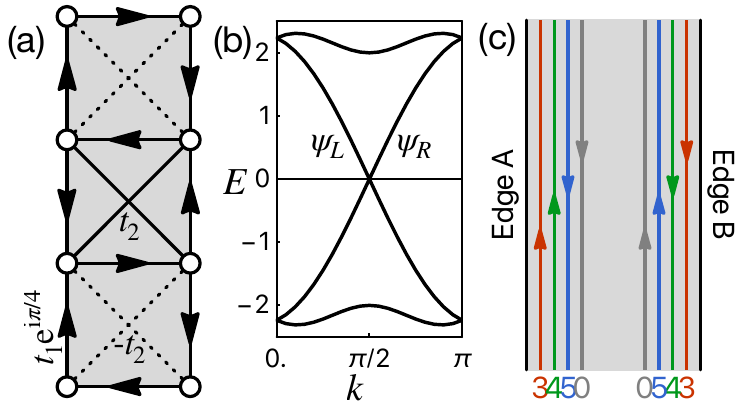}
\caption{(Color online.) (a) The fermion hopping pattern on the two-leg ladder lattice for the first layer. Arrow link: $t_1\mathrm{e}^{\ii \pi/4}$ (along the arrow direction), solid link: $t_2$, dashed link: $-t_2$. This (2+1)D thin strip is effectively the same as (1+1)D by regarding the finite-width dimension as internal degrees of freedom of the (1+1)D system. (b) Energy dispersion for $t_1=1,t_2=0.5$. Gapless edge modes are strictly localized on the two boundaries of the ladder. (c) Schematic diagram showing the configuration of the four flavors of chiral fermions on the edges.}
\label{fig:model}
\end{figure}

Following Wang-Wen's chiral fermion model \cite{Wang2013Non-Perturbative, wang2019solution}, the (1+1)D chiral fermions and their mirror partners can be viewed as the chiral edge modes on the opposite boundaries of a (2+1)D multi-layer Chern insulator \cite{Haldane1988Model}, each layer with a Chern number $\pm 1$. To construct the chiral fermions on a lattice, we start with four layers of Chern insulators on a two-leg ladder as shown in \figref{fig:model}(a). On each lattice site $i$, we introduce four complex fermions, described by the annihilation operators  $\psi_{i,I}$ (with $I=1,2,3,4$ being the layer/flavor index). The fermion hopping is governed by the lattice Hamiltonian
\eq{H_\text{free}=\sum_{I=1}^{4}\sum_{i,j}(t_{I,ij}\psi_{I,i}^\dagger \psi_{I,j}+\text{h.c.}),}
where the hopping parameters $t_{I,ij}$ are non-zero only on the nearest and next-nearest neighbor links. For the first two layers $I=1,2$, the nearest neighbor hoppings are purely imaginary with $t_{I,ij}=\e^{\ii \pi/4} t_1$ if $j\to i$ follows the link direction, and the next-nearest neighbor hoppings are real with $t_{I,ij}=t_2$ (or $-t_2$) on the solid (or dashed) links, as shown in \figref{fig:model}(a). We fix $t_1=1$ and $t_2=0.5$. This hopping pattern ensures a $\pi$ Berry flux through each square plaquette, realizing a minimal model of Chern insulator in each layer. For the last two layers $I=3,4$, the hopping parameters are complex conjugated, such that the band Chern numbers in the last two layers are opposite to those of the first two layers.

The lattice model has a four-site unit cell that repeats along the ladder direction, hence the lattice momentum $k$ along the ladder direction is a good quantum number, and the system is effectively (1+1)D. In each layer, the single-particle energy dispersion (band structure) is shown in \figref{fig:model}(b), which includes two gapped bulk bands together and two gapless  edge modes of opposite velocities (localized separately on the two boundaries). Stacking all layers together, the lattice model realizes four chiral fermions (as two pairs of counter-propagating modes) on each edge, as illustrated in \figref{fig:model}(c). Since the four layers of fermions are decoupled at the free fermion level, we are free to assign them with the 3,4,5,0 chiral $\U(1)$ charges respectively, such that the low-energy edge modes realize the 3-4-5-0 chiral fermions and their mirror partners. We treat the edge A as the light (chiral fermion) sector, and the edge B as the mirror sector (to be gapped out). If we can generate a mass gap for the edge B fermions only without breaking the chiral $\U(1)$ symmetry, we will succeed in achieving a lattice regularization of the 3-4-5-0 field theory \eqnref{eq:3450} in this (1+1)D system in terms of the gapless edge A fermions.

The fact that the $\U(1)$ 't Hooft anomaly vanishes for the 3-4-5-0 model indicates that it should be possible to gap out the edge B fermions trivially without breaking the chiral $\U(1)$ symmetry. However, the chiral $\U(1)$ symmetry is restrictive enough to prevent the gapping to happen on the free-fermion level. 
Therefore, we resort to the idea of gapping out the mirror fermions by interactions, which has been previously explored by Chen, Giedt and Poppitz (CGP) \cite{chen2013decoupling} in the 3-4-5-0 lattice model, where all $\U(1)$ symmetry allowed interactions are included. However, 
the CGP result shows a singular non-local behavior for the gauge field polarization tensor in the mirror sector, which indicates the mirror sector still has surviving gapless modes charged under the gauge field. The reason could be that the CGP approach introduces too many interaction terms, and some of them are harmful. In order to achieve the SMG, the fermion interaction must be carefully selected to satisfy the gapping condition (i.e.~the interaction operators must be self-bosonic and mutual-bosonic in terms of the operator braiding statistics \cite{Haldane1995xgi, Kapustin1008.0654, WangWen1212.4863, Levin1301.7355}), as elaborated in recent works  \cite{Wang2013Non-Perturbative,wang2019solution}. It turns out that the lowest order interactions that satisfy the gapping condition are the following six-fermion local interactions \cite{Wang2013Non-Perturbative},
\eqs{\label{eq:H_int}H_\text{int}=\sum_{i\in\text{B}}g_1&(\psi_{1,i}\psi_{2,i}^\dagger\psi_{2,i+1}^\dagger\psi_{3,i}\psi_{4,i}\psi_{4,i+1}+\text{h.c.})\\
+g_2&(\psi_{1,i}\psi_{1,i+1}\psi_{2,i}\psi_{3,i}^\dagger\psi_{3,i+1}^\dagger\psi_{4,i}+\text{h.c.}),}
which are seemly irrelevant dimension-5 operators in the perturbative RG around the gapless free fermion fixed point. 
The interaction respects the chiral $\U(1)$ symmetry, and is only applied to sites on the B edge (denoted as $i\in \text{B}$), with $i+1$ being the next site of $i$ along the B edge. The interaction looks highly irrelevant in the free-fermion limit. 
However, strong enough interaction (strong in the sense that the interaction energy scale $E_{\rm int}$ is large but still in the same order 
of magnitude as the kinetic energy $E_{\rm free}$, 
thus $E_{\rm int}/E_{\rm free}\simeq \mathcal{O}(1)$ is nonperturbative) 
may still generate non-perturbative effect that gaps out the edge B fermions. Our central goal
is to numerically verify that the proposed interaction \eqnref{eq:H_int} indeed drives the SMG in and only in the mirror sector.

\begin{figure}[]
\includegraphics[width=0.9\linewidth]{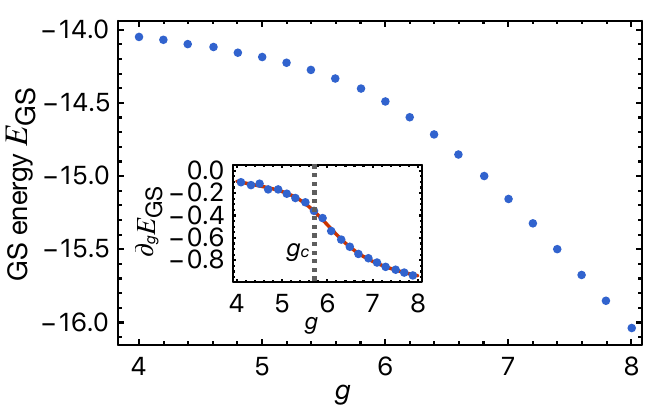}
\caption{(Color online.) Ground state (GS) energy per unit cell as a function of interaction strength $g$. The inset shows the first-order derivative of the GS energy with respect to $g$. The features around $g\approx 5.7$ (indicated by the grey dashed line) signal a quantum phase transition.}
\label{fig2}
\end{figure}

\emph{DMRG results.} --- We study the lattice model $H=H_\text{free}+H_\text{int}$ by the DMRG method \cite{White1992Density} using the ITensor software library \cite{itensor}. For simplicity, we set $g_1=g_2=g$ as the only interaction parameter. The simulation is performed on a two-leg ladder of 20 unit cells, where three different matrix product state (MPS) bond dimensions $\mathcal{D}=6000,7000,8000$ are used (the code to reproduce our result is available at \cite{github}). Computed physical quantities are then extrapolated to the $\mathcal{D}\to\infty$ limit assuming a $1/\mathcal{D}$ scaling. 
Figure~\ref{fig2} shows the ground state energy $E_\text{GS}$ (of the full Hamiltonian $H$) per unit cell as a function of the interaction strength $g$, where the inset shows its first-order derivative $\partial_g E_\text{GS}$. The onset of a non-zero $\partial_g E_\text{GS}=g^{-1}\langle H_\text{int}\rangle$ around $g_c\approx 5.7$ signifies the development of the $\langle H_\text{int}\rangle\neq 0$ condensation across the SMG transition. The smooth kink of $\partial_g E_\text{GS}$ indicates a (high-order) continuous transition (in fact, as to be shown soon, this should be an infinite-order transition with no singularity in $E_\text{GS}$ and its derivatives to any order).

\begin{figure}[]
\includegraphics[width=\linewidth]{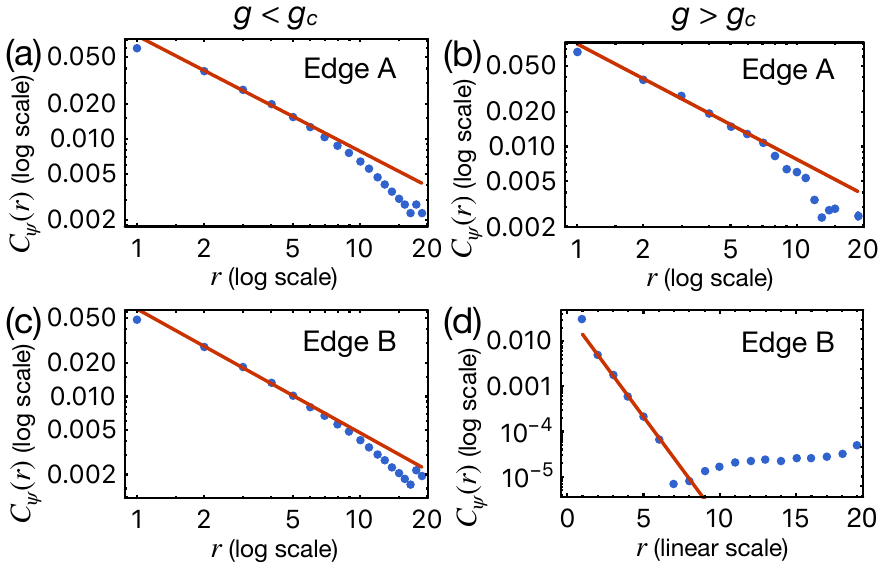}
\caption{(Color online.) Correlations on both edges before and after transition. Linear fit (red line) is performed for intermediate distances from $r=2$ to $r=6$ in each case, in order to faithfully extract the low energy physics while avoiding the artifacts due to the gap caused by finite bond dimension in the matrix product state representation. (a) $g=5.0<g_c$ for edge A. The log-log plot shows a power-law decay for intermediate distances. (b) $g=7.0>g_c$ for edge A. The log-log plot again shows a power-law decay. (c) $g=5.0<g_c$ for edge B. The log-log plot shows a power-law decay. (d) $g=7.0>g_c$ for edge B. The semi-log plot indicates an exponential decay, i.e. edge B becomes gapped.}
\label{fig3}
\end{figure}

To further confirm the existence of the critical point $g_c$, we calculate the fermion correlation functions $C_\psi(r) \equiv \langle\psi_{I,i+r}^\dagger \psi_{I,i}\rangle$ on both edge A and edge B across the transition. It turns out that the behavior of $C_\psi$ is the same for all $I=1,2,3,4$, such that it is sufficient to show one of the  four flavors. 
Figures~\ref{fig3} (a,c) and Figs.~\ref{fig3} (b,d) show the correlation functions for each edge before and after the transition, respectively. We observe that edge A is always gapless with power-law correlations. In contrast, edge B is gapless when $g<g_c$ but becomes gapped with an exponential-decay correlation when $g>g_c$. The two qualitatively different behaviors must be separated by a quantum phase transition.

\begin{figure}[]
\includegraphics[width=0.9\linewidth]{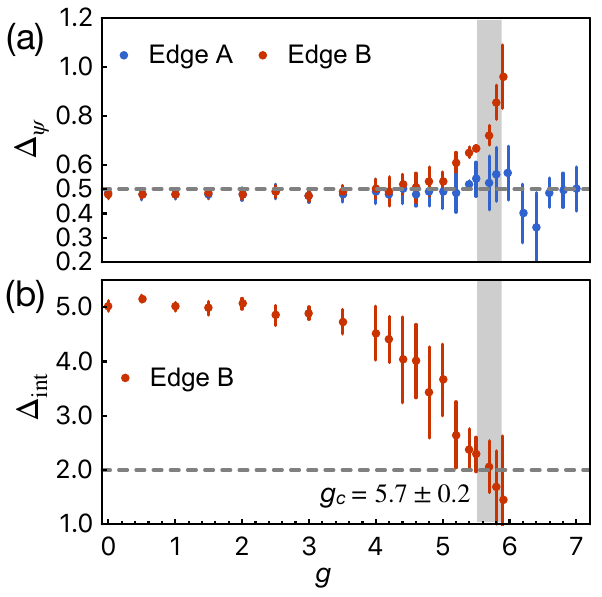}
\caption{(Color online.) (a) The evolution of fermion scaling dimension $\Delta_\psi$ on both edges as the interaction strength $g$ approaches the critical point. The scaling dimension is obtained from the power-law  fitting as in Fig.~\ref{fig3}. The horizontal dashed line indicates the free fermion limit. The gray stripe shows the estimated critical interaction strength $g_c$ with some uncertainty. (b) The solved scaling dimension for the interaction terms on edge B based on the scaling dimensions of multiple operators (Refer to Appendix~\ref{append::RG-scaling-dim} for details). The horizontal dashed line indicates the marginal value 2 of $\Delta_{\text{int}}$, across which the phase transition is expected to happen.}
\label{fig4}
\end{figure}

By fitting the power-law correlation function $C_\psi(r)\sim 1/r^{2\Delta_\psi}$ before the transition, we can extract the fermion scaling dimension $\Delta_\psi$ on both edges. The result is shown in \figref{fig4}(a). In the free-fermion limit ($g=0$), the fermion scaling dimension is $\Delta_\psi=\frac{1}{2}$ on both edges. The finite-size effect tends to reduce the scaling dimension slightly. A finite-size scaling of the scaling dimension in the free fermion limit is performed in Appendix~\ref{append::free-fermion-scaling-dim}, confirming that our result converges to the long-distance limit correctly. As $g$ increases towards $g_c$, the fermion scaling dimension on the edge B increases continuously from $\frac{1}{2}$ to about $0.67$ (near $g_c$), indicating that fermion operators get renormalized by the interaction significantly. For $g>g_c$, the correlation on  the edge B becomes short-ranged, such that the fermion scaling dimension is no longer defined (although the power-law fitting on the finite-size data will continue give some estimated exponent that extrapolates beyond the critical point before the correlation length shrinks below the  system size). However, on the edge A, the fermion scaling dimension, while experiencing some fluctuations near the critical point, generally stays close to the free fermion limit regardless of the interaction strength. The scaling dimension remains stable even after $g$ goes across the transition point $g_c$ by a significant amount. This implies that the edge A remains gapless and almost free, as the edge B interaction can only induce a perturbative interaction on the edge A through the proximity effect.

To verify that the chiral $\U(1)$ symmetry is not broken spontaneously by the condensation of fermion bilinear masses, we measure correlation functions of Dirac and Majorana mass operators on the B edge, i.e.~$C_{\psi_I^\dagger\psi_J}(r) \equiv \langle \psi_{J,i+r}^\dagger \psi_{I,i+r}\psi_{I,i}^\dagger \psi_{J,i}\rangle$ and $C_{\psi_I\psi_J}(r) \equiv \langle \psi_{J,i+r}^\dagger \psi_{I,i+r}^\dagger \psi_{I,i} \psi_{J,i}\rangle$. \figref{fig2-append} in the Appendix~\ref{append::mass-correlations} shows the correlations for all the eight mass terms are short-ranged (exponential decay) along the B edge in the strong coupling gapped phase ($g>g_c$), which confirms that the mirror fermions on the B edge are gapped by the SMG mechanism without long-range ordering of bilinear masses. Therefore, the remaining gapless fermions on the A edge successfully realize the lattice regularization of chiral fermions in the 3-4-5-0 model preserving the chiral $\U(1)$ symmetry.


\emph{Luttinger liquid RG analysis.} --- To better understand the nature of the SMG transition at $g_c$, we perform the Luttinger liquid RG analysis for the edge B fermions. We first bosonize the mirror fermions by $\psi_I\sim \e^{\ii\varphi_I}$. Then the (1+1)D interacting mirror fermions can be described by the Luttinger liquid effective field theory in terms of the $\varphi=(\varphi_1,\varphi_2,\varphi_3,\varphi_4)^\intercal$ fields
\eq{\label{eq:LL}\scL=\frac{1}{4\pi}(
\partial_t\varphi^\intercal K \partial_x\varphi+\partial_x\varphi^\intercal V \partial_x\varphi)+\sum_{\alpha=1,2}g_\alpha\cos(l_\alpha^\intercal \varphi),}
where $K=\sigma^{30}$ and $V=\sigma^{00}$ (in the UV limit) are $4\times 4$ matrixes (where $\sigma^{\mu\nu}=\sigma^\mu\otimes\sigma^\nu$ denotes the tensor product of Pauli matrices). The two interaction terms $g_1,g_2$ in \eqnref{eq:H_int} correspond to the cosine-terms in \eqnref{eq:LL} specified by the vectors $l_1=(1,-2,1,2)^\intercal$ and $l_2=(2,1,-2,1)^\intercal$, respectively. The RG flow with respect to the log-energy-scale $\ell=-\ln \Lambda$ is given by \cite{Katzir2020Superconducting}
\eqs{\frac{\dd g_\alpha}{\dd \ell}&=(2-\Delta_{l_\alpha})g_\alpha-\frac{1}{2}\sum_{l_\beta\pm l_\gamma=l_\alpha}g_\beta g_\gamma,\\
\frac{\dd V^{-1}}{\dd\ell}&=\frac{1}{2}\sum_{\alpha}g_\alpha^2(K^{-1}l_\alpha l_\alpha^\intercal K^{-1}-V^{-1}l_\alpha l_\alpha^\intercal V^{-1}),}
where $\Delta_l=\frac{1}{2}l^\intercal V^{-1} l$ denotes the scaling dimension of the vertex operator $\e^{\ii l^\intercal \varphi}$. Under the RG flow, the $V$ matrix gets renormalized to the general form
\eq{V=\sqrt{1+y_1^2+y_2^2}\;\sigma^{00}-y_1\,\sigma^{10}-y_2\,\sigma^{22},}
where $y_1,y_2$ are Luttinger parameters that depend on the RG scale $\ell$. \tabref{tab:scaling} concludes the analytical expressions of the scaling dimensions $\Delta_l$ of the fermion, Dirac mass and Majorana mass operators given their corresponding charge vectors $l$. We numerically determine the scaling dimensions of these operators before the transition $(g<g_c)$, by fitting the power-law exponents of their correlation functions (see Appendix.~\ref{append::RG-scaling-dim} for details). 

\begin{table}[htp]
\caption{Operator scaling dimensions and Luttinger parameters at the free-fermion limit ($g=0$) and at the critical point ($g=g_c$).}
\begin{center}
\begin{tabular}{lcc|cc}
 & $l^\intercal$ & $\Delta_l$ & free & critical\\
\hline
$\psi_1$ & $(1,0,0,0)$ & $\frac{1}{2}\sqrt{1+y_1^2+y_2^2}$ & $\frac{1}{2}$ & $0.67\pm 0.07$\\
$\psi_1^\dagger\psi_3$ & $(-1,0,1,0)$ & $\sqrt{1+y_1^2+y_2^2}-y_1$ & $1$ & $0.76\pm 0.05$\\
$\psi_1\psi_4$ & $(1,0,0,1)$ & $\sqrt{1+y_1^2+y_2^2}-y_2$ & $1$ & $0.73\pm 0.01$\\
\hline
& & $y_1$ & 0 & $0.72\pm 0.18$\\
& & $y_2$ & 0 & $0.70\pm 0.15$\\
\end{tabular}
\end{center}
\label{tab:scaling}
\end{table}%

From the scaling dimensions, we infer the Luttinger parameters $y_1,y_2$, and calculate the scaling dimension of the interaction operator $\Delta_\text{int}:=\Delta_{l_1}=\Delta_{l_2}=\sqrt{1+y_1^2+y_2^2}-3y_1-4y_2$. The evolution of $\Delta_\text{int}$ is shown in \figref{fig4}(b), which drops continuously from $\Delta_\text{int}=5$ at the free-fermion limit ($g=0$) to $2.17\pm 0.27$ at the SMG transition ($g=g_c$). Although the interaction is perturbatively irrelevant at the free-fermion fixed point, finite strength of the interaction can renormalize the Luttinger parameter, which reduces its own scaling dimension. Our numerical result indicates that the SMG transition is triggered exactly when the interaction scaling dimension is reduced to marginal $\Delta_\text{int}=2$, which matches the mechanism of the Berezinskii-Kosterlitz-Thouless (BKT) transition. This scenario was also proposed by Tong in a recent theoretical study \cite{Tong2021Comments}. Our numerical study provides more detailed RG analysis and more solid evidence in support of the BKT transition scenario.

\emph{Conclusion and Discussions.} --- We numerically demonstrate the lattice regularization of (1+1)D chiral fermions in the 3-4-5-0 model. This is achieved by gapping out the anomaly-free mirror sector using properly designed interactions via the SMG mechanism, leaving the light sector gapless. By simulating the lattice model with the DMRG method, we identify the SMG transition point $g_c$. In the strong coupling phase ($g=g_c$), we show that the mirror fermions are gapped without breaking the chiral symmetry, and the light fermions remain gapless. We numerically determine the scaling dimension of the interaction operator before the transition, which evolves continuously from irrelevant to marginal. This behavior clearly indicates the BKT nature of the SMG transition in our model. 
Once the anomaly-free U(1) symmetry is dynamically gauged, we expect to obtain
a (1+1)D lattice chiral gauge theory coupled to chiral fermions, which could potentially be simulated by the quantum Monte Carlo method \cite{Chen2013Model345}, as our proposed six-fermion interaction in \eqnref{eq:H_int} admits the following Yukawa decomposition (with site indices omitted for brevity)
\eqs{H_\text{Yuk}=&(\phi_1^2\psi_1\psi_3+\phi_1^\dagger\psi_2^\dagger \psi_4+\text{h.c.})+\frac{1}{\tilde{g}_1}\phi_1^\dagger\phi_1\\
+&(\phi_2^2\psi_2\psi_4+\phi_2^\dagger\psi_1\psi_3^\dagger+\text{h.c.})+\frac{1}{\tilde{g}_2}\phi_2^\dagger\phi_2,}
such that integrating out the Yukawa bosons $\phi_\alpha$ reproduces our interaction at the leading order of $g_\alpha\sim \tilde{g}_\alpha^2$. 
Based on the equivalence between 
the U(1) anomaly-free and gapping conditions in (1+1)D \cite{Wang2013Non-Perturbative, wang2019solution}, hopefully our work can prompt
future simulations on other (1+1)D lattice chiral fermion/gauge theory models.

\textit{Acknowledgement.} --- MZ would like to thank Miles Stoudenmire for helpful discussions on using the ITensor package. ZZ is supported by the National Natural Science Foundation of China (Grant No. 12074375) and the start-up funding of KITS at UCAS (Grant No.118900M026). JW is supported by
Harvard CMSA. MZ and YZY are supported by a start-up funding of UCSD.

\bibliographystyle{apsrev4-2}
\bibliography{ref}%

\appendix
\section{Scaling dimensions in free fermion limit}
\label{append::free-fermion-scaling-dim}
In the free fermion limit, correlation functions and consequently scaling dimensions can be calculated analytically. Here we work out the free fermion case on the lattice as a benchmark to the DMRG calculation at $g=0$.
\subsection{Fermion scaling dimension}
\begin{figure}[]
\includegraphics[width=0.9\linewidth]{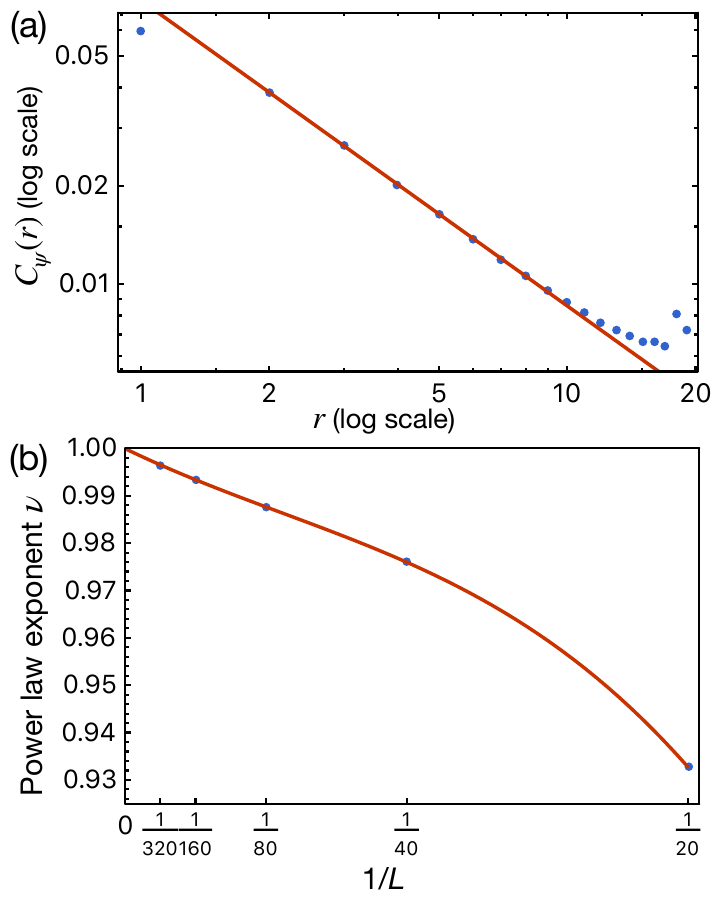}
\caption{(a) Linear fit for the correlation function on a log-log scale when the system has 20 unit cells. The power law exponent obtained is around 0.936, which is smaller than 1. (b) Finite-size scaling for the exponent using a polynomial function for system sizes $L=20,40,80,160,320$. The extrapolation to $L=\infty$ recovers the ideal $\nu=1$ limit.}
\label{fig1-append}
\end{figure}

For the free fermion-fermion correlation, we can just pick the fermion flavor $f_3$ without loss of generality. The two-point correlation is given by (with the flavor index ignored)
\begin{equation}
    C_\psi(r)=\langle \psi_{i+r}^\dagger \psi_i\rangle \sim r^{-\nu}.
\end{equation}
On a finite lattice with real space Hamiltonian $\mathcal{H}$, we can do a change basis from the fermionic operators in real space fermionic operators in energy eigenspace, so that the correlation at half-filling can be more conveniently calculated. Assuming $\mathcal{H}$ has eigenstates $|\epsilon_n\rangle$ with corresponding eigenenergies $\epsilon_n$, then the change of basis is given by $\psi^\dagger(i)=\sum_n\langle \epsilon_n|i\rangle \psi_n^\dagger$. Eventually the correlation becomes
\begin{equation}\label{eq::corr-free}
    C_\psi(r)=\sum_{n,m}\langle \epsilon_n|i\rangle\langle j|\epsilon_m\rangle\langle \psi_n^\dagger \psi_m\rangle.
\end{equation}
Since the ground state is half-filled, the above summation in $n$ or $m$ is only over the lower half of the energy spectrum. The system we have in the main text consists of 20 unit cells. Exact diagonalization can be done in the free fermion limit and the correlation can then be calculated using Eq.~(\ref{eq::corr-free}). With open boundary condition (same as the DMRG setup), the correlation with log-log scale is plotted in Fig.~\ref{fig1-append}(a) together with a linear fit. The power law exponent obtained is less than 1, which explains the deviation of the scaling dimension $\Delta_\psi$ from 0.5 for $g=0$ in Fig.~\ref{fig4}. This deviation is mainly a finite-size effect, demonstrated in Fig.~\ref{fig1-append}(b). In the large-system-size limit, the free fermion scaling dimension of 0.5 is recovered.

\subsection{Mass term scaling dimensions}
Other than the fermion-fermion correlation, correlations of various mass terms on a finite lattice can also be calculated analytically in the free fermion limit. In this case, the expected scaling dimension for a bilinear in the large-system-size limit is 1.  Taking the fermion bilinear $\psi_1\psi_3^\dagger$ as an example, the correlation on the lattice is given by (using Wick's theorem)
\begin{equation}
\begin{split}
    C_{\psi_1^\dagger\psi_3}(r)&=\langle \psi_{1,i+r}^\dagger\psi_{3,i+r}\psi_{3,i}^\dagger\psi_{1,i}\rangle\\
    &=\langle \psi_{1,i+r}^\dagger\psi_{1,i}\rangle \langle\psi_{3,i+r}\psi_{3,i}^\dagger\rangle\\
    &\sim r^{-2\nu},
\end{split}
\end{equation}
i.e., the fermion bilinear exponent is simply double of that for the single fermion, as expected. Therefore, the finite-size behavior should also be the same.
\section{Mass term correlations in the gapped phase for edge B}
\label{append::mass-correlations}
\begin{figure}[]
\includegraphics[width=0.9\linewidth]{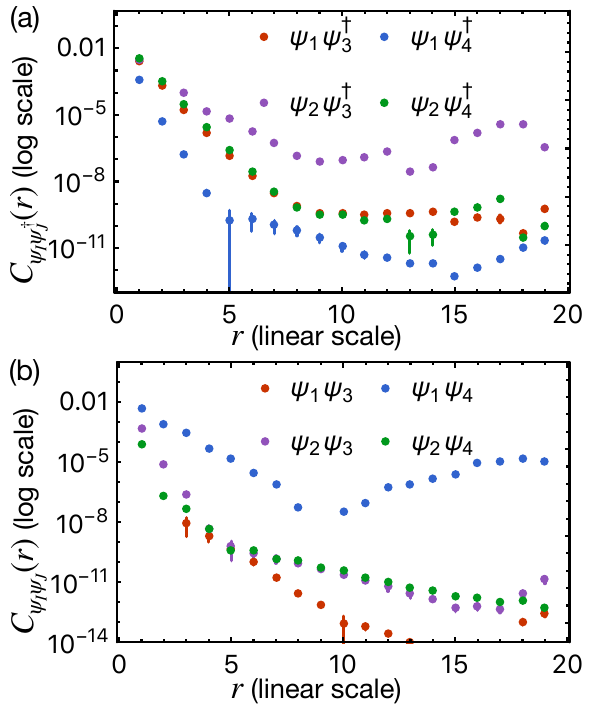}
\caption{Semi-log correlations for the various mass terms in the gapped phase. (a) Dirac masses $\psi_I\psi_J^\dagger$. The interaction strengths chosen for the four mass terms $\psi_1\psi_3^\dagger$, $\psi_1\psi_4^\dagger$, $\psi_2\psi_3^\dagger$ and $\psi_2\psi_4^\dagger$ in the gapped phase are $6.5$, $6.5$, $6.6$ and $6.4$ respectively. (b) Majorana masses $\psi_I\psi_J$. The interaction strengths chosen for the four mass terms $\psi_1\psi_3$, $\psi_1\psi_4$, $\psi_2\psi_3$ and $\psi_2\psi_4$ in the gapped phase are $6.4$, $6.3$, $6.3$ and $7.2$ respectively.}
\label{fig2-append}
\end{figure}

In this section, we present the correlations for the mass terms on edge B after the gapping transition to demonstrate that the U(1) chiral symmetry is preserved in the gapped phase. The correlations for the Dirac masses are shown in \figref{fig2-append}(a) and the correlations for the Majorana masses are shown in \figref{fig2-append}(b). We see clear evidence for exponential decays for all the eight different mass terms at relatively shorter length scales $r\lesssim 5$, where the correlation is expected to be dominated by the SMG gap. The non-monotonic behavior of the correlation function for larger distance $r \gtrsim 5$ is an artifact arising from the finite MPS bond dimension, and should not be trusted. Different interaction strengths are chosen for the different mass terms in order to better demonstrate the exponential decay features. Thus, we conclude that the chiral U(1) symmetry is preserved in the gapped phase on edge B. 
\section{Scaling dimensions under RG}
\label{append::RG-scaling-dim}
The scaling dimensions of some of the interesting operators are summarized in Table~\ref{table::scaling-dim}.

\begin{table}[!htbp]
	\begin{ruledtabular}
		\begin{tabular}{c|c|c}
	    Operators & $l$ & $\Delta_l$\\ 
	    \hline
	    $\psi_1$ & $(1,0,0,0)$ & $\frac{1}{2}\sqrt{1+y_1^2+y_2^2}$\\
	    \hline
	    $\psi_1\psi_3^\dagger$ & $(1,0,-1,0)$ & $-y_1+\sqrt{1+y_1^2+y_2^2}$\\
	    \hline
	     $\psi_1\psi_4^\dagger$ & $(1,0,0,-1)$ & $y_2+\sqrt{1+y_1^2+y_2^2}$\\
	    \hline
	     $\psi_2\psi_3^\dagger$ & $(0,1,-1,0)$ & $-y_2+\sqrt{1+y_1^2+y_2^2}$\\
	    \hline
	     $\psi_2\psi_4^\dagger$ & $(0,1,0,-1)$ & $-y_1+\sqrt{1+y_1^2+y_2^2}$\\
	    \hline
		 $\psi_1\psi_3$ & $(1,0,1,0)$ & $y_1+\sqrt{1+y_1^2+y_2^2}$\\
	    \hline	
	     $\psi_1\psi_4$ & $(1,0,0,1)$ & $-y_2+\sqrt{1+y_1^2+y_2^2}$\\
	    \hline	
	     $\psi_2\psi_3$ & $(0,1,1,0)$ & $y_2+\sqrt{1+y_1^2+y_2^2}$\\
	    \hline	
	     $\psi_2\psi_4$ & $(0,1,0,1)$ & $y_1+\sqrt{1+y_1^2+y_2^2}$\\
	    \hline
	    $\psi_1\psi_3\psi_2^\dagger\partial_x\psi_2^\dagger\psi_4\partial_x\psi_4$ & $(1,-2,1,2)$& $-3y_1-4y_2+5\sqrt{1+y_1^2+y_2^2}$\\
	    \hline
	    $\psi_1\partial_x\psi_1\psi_3^\dagger\partial_x\psi_3^\dagger\psi_2\psi_4$ &$(2,1,-2,1)$ & $
	    -3y_1-4y_2+5\sqrt{1+y_1^2+y_2^2}$\\
	    
		\end{tabular}
		\caption{\label{table::scaling-dim}Scaling dimensions of the various operators, some of which are used to solve the two parameters $y_1$ and $y_2$.}
	\end{ruledtabular}
\end{table}

\begin{figure}[]
\includegraphics[width=0.9\linewidth]{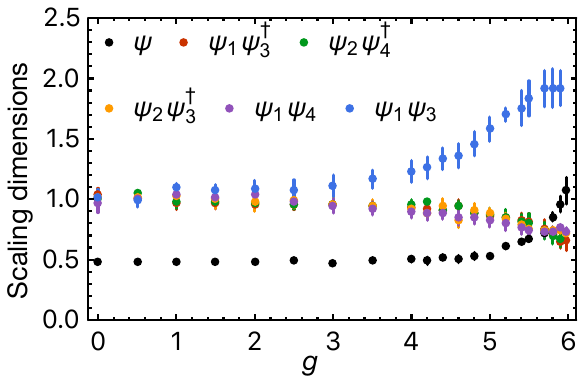}
\caption{Evolution of scaling dimensions for some of the mass terms under RG on edge B. The fermion scaling dimension $\Delta_\psi$ is also shown for comparison.}
\label{fig3-append}
\end{figure}

These scaling dimensions can in principle be measured by calculating the power-law correlation using DMRG. Here we are interested in the scaling dimensions of the gapping terms under RG before the Luttinger liquid becomes gapped. However, for large scaling dimensions, the power-law decay is too fast to be measured accurately. Therefore, in order to solve for the parameters $y_1$ and $y_2$, we only use the correlations for operators with relatively smaller scaling dimensions, which, based on Table~\ref{table::scaling-dim}, include $\psi_1,\psi_1\psi_3^\dagger,\psi_2\psi_3^\dagger,\psi_2\psi_4^\dagger$ and $\psi_1\psi_4$. Fig.~\ref{fig3-append} shows the evolution of the scaling dimensions for some of the mass terms under RG flow.

\begin{figure}[]
\includegraphics[width=0.9\linewidth]{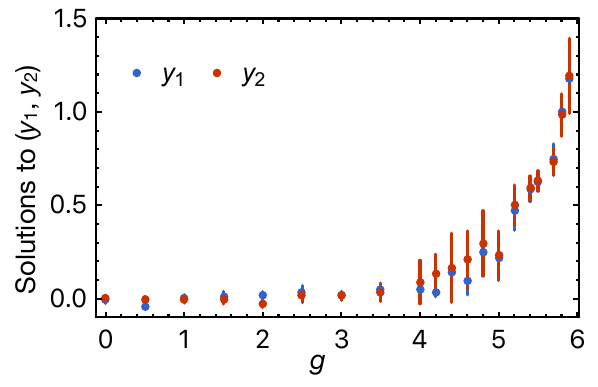}
\caption{Solutions to the Luttinger parameters $(y_1,y_2)$ based on RG equation and some of the measured scaling dimensions using DMRG.}
\label{fig4-append}
\end{figure}

Instead of using two of the different scaling dimensions to solve for the two parameters, we try to make use of all the five aforementioned operators to have a faithful representation of the available DMRG data. For a particular interaction strength $g$, denote the five scaling dimensions from RG calculation by $\Delta_i^{(\text{RG})}(y_1(g),y_2(g))$ and the DMRG counterparts by $\Delta_i^{(\text{DMRG})}(g)$ with errors $\delta\Delta_i^{(\text{DMRG})}(g)$. Then $y_1$ and $y_2$ are solved by minimizing the following error function at each $g$:
\begin{equation}
    f_{\text{error}}(g)=\sum_i\frac{\left(\Delta_i^{(\text{RG})}(y_1(g),y_2(g))-\Delta_i^{(\text{DMRG})}(g)\right)^2}{\delta\Delta_i^{(\text{DMRG})}(g)^2},
\end{equation}
where $\frac{1}{\delta\Delta_i^{(\text{DMRG})}(g)^2}$ can be considered as the weight of each contribution to the total error function. The solutions $(y_1,y_2)$ with error bars are obtained in the following way. At each $g$, a numerical value for the DMRG scaling dimension is randomly drawn from the interval $[\Delta_i^{(\text{DMRG})}-\delta\Delta_i^{(\text{DMRG})},\Delta_i^{(\text{DMRG})}+\delta\Delta_i^{(\text{DMRG})}]$, then the error function $f_{\text{error}}$ is minimized to find $(y_1,y_2)$, with the weight of this particular solution given by $\frac{1}{f_{\text{error}}^2}$. This process is repeated 100 times for each $g$, after which weighted average is taken to obtain the mean of $(y_1,y_2)$ with the errors given by the weighted uncertainty. The evolution of the two parameters with RG flow is shown in Fig.~\ref{fig4-append}. The solved $y_1$ and $y_2$ can then be used to calculate the scaling dimension of the gapping terms $\Delta_{\text{int}}$ plotted in Fig.~\ref{fig4}(b) in the main text.

\end{document}